\documentclass[%
 reprint,
superscriptaddress,
 amsmath,amssymb,
 aps,
]{revtex4-1}

\usepackage{graphicx}
\usepackage{dcolumn}
\usepackage{bm}

\begin{document}

\newcommand{\mvec}[2]
{
\left(\begin{array}{c}
#1  \\
#2  
\end{array}
\right)
}

\newcommand{\mmat}[4]
{
\left(\begin{array}{cc}
#1  & #2\\
#3  & #4
\end{array}
\right)
}


\title{Generalized Chiral Symmetry and Stability of Zero Modes\\
for Tilted Dirac Cones}

\author{Tohru Kawarabayashi}
\affiliation{Department of Physics, Toho University,
Funabashi, 274-8510 Japan}

\author{Yasuhiro Hatsugai}
\affiliation{Institute of Physics, University of Tsukuba, Tsukuba, 305-8571 Japan}

\author{Takahiro Morimoto}
\affiliation{Department of Physics, University of Tokyo, Hongo, 
Tokyo 113-0033 Japan }

\author{Hideo Aoki}
\affiliation{Department of Physics, University of Tokyo, Hongo, 
Tokyo 113-0033 Japan }

\date{\today}

\begin{abstract}
While it has been well-known that the chirality is an important symmetry 
for Dirac-fermion systems that gives rise to the zero-mode Landau level in graphene, here we explore whether this notion can be extended to tilted Dirac cones 
as encountered in  organic metals.  
We have found that there exists 
a ''generalized chiral symmetry" that encompasses the tilted Dirac cones, 
where a generalized 
chiral operator $\gamma$, satisfying 
$\gamma^{\dagger} H + H\gamma =0$ 
for the Hamiltonian $H$, protects the zero mode. 
We can use this to show that the $n=0$ Landau level is delta-function-like 
(with no broadening) by extending the Aharonov-Casher argument. 
We have numerically confirmed that 
a lattice model that possesses the generalized chirality 
has an anomalously sharp Landau level for 
spatially correlated randomness.  
\end{abstract}

\pacs{73.43.-f, 72.10.-d, 71.23.-k}

\maketitle

{\it Introduction ---} 
Since the seminal observation of the quantum Hall effect in graphene,\cite{Geim,Kim}
the zero modes of the massless Dirac fermion, which is essential for the characteristic quantum Hall effect in graphene,  
has been intensively discussed.  
Specifically, the zero mode is stable against ripples in graphene, 
which has been discussed in terms of the index theorem, or more 
explicitly for wave functions with the argument due to 
Aharonov and Chasher \cite{Geim,AC,KN,Neto,Kail,VKG,Hatsu10}.  For the stability of zero modes, a crucial ingredient is the 
chiral symmetry defined as the existence of an operator $\Gamma$ that anti-commutes
with 
the effective Hamiltonian $H$, $\{ \Gamma ,H\} = 0$, with $\Gamma^2=1$, which is more relevant than the Dirac-cone dispersion per se, 
since the stability is also 
shown for a chiral symmetric bilayer graphene with a quadratic dispersion \cite{MF,KP}. 
In the case of graphene, the Dirac cone is vertical, so that 
the chiral operator has a simplest possible form of $\Gamma=\sigma_z$.  
Even when the system is disordered, the 
anomalous criticality at the $n=0$ Landau level is retained 
as far as the randomness  respects the chiral symmetry \cite{LFSG,OGM,Guinea,KHA}. 
The stability of zero modes
has been observed experimentally as well for a mono-layer graphene \cite{GZKPMM}.

On the other hand, we encounter Dirac cones in a wider class of 
materials, where the effective theory is 
more generally described by {\it tilted Dirac cones}, as 
is the case with an organic material $\alpha$-(BEDT-TTF)$_2$I$_3$ \cite{KKS,KKSF,KSFG,KM,TSTNK,TSKNK,STTKSNK}.   
Since the conventional chiral symmetry 
is broken in tilted  Dirac cones, it becomes an essential question to 
ask whether (i) the symmetry is entirely broken, and 
(ii) whether the anomaly in the systems with usual chiral symmetry 
is washed out in tilted Dirac cones.  
In the absence of disorder, the eigenvalues and 
eigenfunctions for tilted Dirac cones in magnetic fields 
have already been obtained in existing literatures \cite{MHT,MT,GFMP}, 
which indicate that the zero-energy modes themselves persist in 
clean tilted cones. 

Effects of disorder on the zero modes is indeed an important issue, 
since, for graphene with the conventional chiral symmtry, it has been 
established that the disorder has an anomalous effect of 
retaining a sharp (delta-function-like) zero Landau level 
accompanied by a sharp (step-function-like) quantum Hall step, if 
the disorder respects the chiral symmetry.  
We can thus pose a question: what is the effect of 
disorder for the zero Landau level in tilted Dirac cones, where 
the conventional chiral symmetry is absent.  
This is exactly the motivation of the present work.  
In particular, we shall look at how the stability of the 
zero modes possessing an 
anomalous criticality at the $n=0$ Landau level  in 
the presence of disorder (e.g., random components in magnetic fields) 
is affected by the breakdown of the conventional chiral symmetry 
in tilted Dirac cones.   Curiously, 
the study leads us to find 
a ``generalized chiral symmetry", which is then shown to give rise to 
an even wider stability of 
the zero modes persisting in tilted Dirac cones.

Thus we shall first generalize the conventional chiral symmetry so that the tilted and untilted Dirac dispersions can be 
captured in a unified manner. 
The generalized chiral symmetry is defined by the existence of an operator $\gamma$, which  is not
necessarily Hermitian, satisfying the relation $\gamma^\dagger H \gamma = -H$ and $\gamma^2=1$.  
This is consistent with the condition for the Dirac operator to be
elliptic that is required for the index theorem \cite{Nakahara}.
It enables us to show a topological stability of 
zero modes that  is present in  tilted Dirac cones.  
With this generalized chiral symmetry, 
we reformulate the eigenvalue problem so that 
the Aharonov-Casher argument \cite{AC} 
for counting the number of zero modes is extended to tilted cones.  
This implies the zero Landau level is indeed delta-function-like.  
We then numerically confirm how these field theoretic treatments 
on the stability of zero modes appears in a lattice fermion 
model that has tilted Dirac cones where the titling is varied continuously.

{\it Formalism ---} 
The effective Hamiltonian  for a tilted Dirac cone 
can be generically expressed as \cite{MHT,MT,GFMP,Hatsu10}
\begin{equation}
 H = \sigma_0(\mbox{\boldmath $W$}\cdot \mbox{\boldmath $\pi$}/\hbar)  + (\mbox{\boldmath $\sigma$} \cdot \mbox{\boldmath $X$})\pi_x/\hbar
 + (\mbox{\boldmath $\sigma$} \cdot \mbox{\boldmath $Y$})\pi_y/\hbar,
 \label{eff-hamiltonian}
\end{equation}
where  $(\sigma_x, \sigma_y, \sigma_z)$ 
 are Pauli matrices while 
 $\sigma_0 $ is a $2\times 2$ unit matrix, 
and $\bm{X},\bm{Y},\bm{W}$ are real.  
The Dirac cone is tilted when ${^t}\bm{W}={(W_x,W_y)}$ is nonzero, while 
${^t}\bm{X}=(X_x,X_y,X_z)$ and ${^t}\bm{Y}=(Y_x,Y_y,Y_z)$ 
characterize the anisotropy of the Dirac cone.  
Here $\bm{\pi}=\bm{p}+e\bm{A}$ $(e>0)$ is 
the dynamical momentum with 
${^t}\bm{p}= -i\hbar(\partial _x,\partial _y)$ and the vector potential 
${^t }\bm{A}= (A_x,A_y)$ 
for the magnetic field $B=\partial _x A_y-\partial _y A_x$
perpendicular to the $x$-$y$ plane.
 Note that the dynamical momentum satisfies the commutation relation
 $[\pi_x, \pi_y] = -i\hbar eB$.
For the conventional Dirac cone ($\bm{W}= 0$), the Hamiltonian has the 
chiral symmetry associated with an operator defined 
(and generalized to anisotropic cases) by 
$\Gamma = \bm{\sigma}\cdot (\bm{X}\times \bm{Y})/|\bm{X}\times 
\bm{Y}|$, which anti-commutes with  
the Hamiltonian $H=(\bm{\sigma} \cdot \bm{X})\pi_x/\hbar
 + (\bm{\sigma} \cdot \bm{Y})\pi_y/\hbar$ with $\Gamma^2 = \sigma_0$\cite{Hatsu10}. 

When the Dirac cone is tilted (with $H$ containing $\sigma_0$ for 
$\bm{W}\neq {\bf 0}$), the conventional chiral symmetry is broken.
However, here we find that a generalized symmetry does 
exist, which can be seen if we introduce a new operator $\gamma$
defined by 
$$
\gamma =  \bm{\sigma}\cdot [(\bm{X}\times \bm{Y}) -i(W_y
 \bm{X} -  W_x\bm{Y})]/\Delta
 $$ 
 with 
 $\Delta^2 = |\bm{X}\times 
\bm{Y}|^2 - (W_y \bm{X} -W_x\bm{Y})^2$.
This operator $\gamma$ is non-Hermitian for $\bm{W}\neq {\bf 0}$, 
but its eigenvalues are $\pm 1$ since $\gamma^2=(\gamma^{\dagger})^2 = \sigma_0$.
We can show that, if $\Delta^2 > 0$, 
$\gamma$ satisfies a relation with the Hamiltonian,
$$
  \gamma^{\dagger} H \gamma = -H,
$$
which we call the {\it generalized chiral symmetry}.
This symmetry reduces to the conventional chiral symmetry 
($\gamma \rightarrow \Gamma$) for $\bm{W}= 0$. 
It should be noted that the cross section of the tilted Dirac cone with a constant energy plane
is an {\it ellipse} as long as $\Delta^2 >0$ (while a hyperbola when $\Delta^2 <0$).

Can we say anything about the wave functions 
as a direct consequence of this generalized chiral symmetry?  
For this purpose  it is instructive to 
choose the (right-) eigenvectors $|\pm \rangle$ of the opeator $\gamma$ 
(with $\gamma |\pm\rangle = \pm |\pm \rangle)$ in the spinor space as a basis of the 
eigenvalue problem for $H$. If we express the (normalized) wave function as 
$
\psi = |+\rangle \psi_+  +  |-\rangle \psi_-, 
$
the Schr\"{o}dinger equation $H \psi = E \psi $ reduces to 
\begin{eqnarray*}
\lefteqn{
\left[\begin{array}{cc}
 \langle + | H | + \rangle & \langle + | H | - \rangle \\
 \langle - | H | + \rangle & \langle - | H | - \rangle
\end{array}\right]
\left[\begin{array}{c}
\psi_+ \\
\psi_-
\end{array}\right] } \nonumber \\
& & \quad \quad =E\left[\begin{array}{cc}
1 & \langle + | - \rangle \\
 \langle - |  + \rangle &1
\end{array}\right] 
\left[\begin{array}{c}
\psi_+ \\
\psi_-
\end{array}\right].
\end{eqnarray*}

Note that the operator $\gamma$, being non-Hermitian, 
has eigenvectors that are in general not orthogonal with each other 
with $\beta = \langle + |-\rangle \neq 0$.
When the generalized chiral symmetry, $\gamma^\dagger H \gamma = -H$, holds, we have $ \langle + | H | + \rangle =  \langle - | H | - \rangle=0$.
The Schr\"{o}dinger equation then becomes 
$$ 
\left[\begin{array}{cc}
0 & \bm{\alpha}\cdot \bm{\pi} \\
\bm{\alpha}^*\cdot \bm{\pi} & 0
\end{array}\right]
\left[\begin{array}{c}
\psi_+ \\
\psi_-
\end{array}\right] 
 =E \left[\begin{array}{cc}
1 & \beta \\
 \beta^* &1
\end{array}\right] 
\left[\begin{array}{c}
\psi_+ \\
\psi_-
\end{array}\right],
$$
where we have introduced a complex 
${^t}\bm{\alpha} = (\alpha_X,\alpha_Y) 
\equiv \hbar^{-1}(\langle + | W_x\sigma_0 + \bm{X}\cdot \bm{\sigma}  |-\rangle, 
\langle + | W_y\sigma_0 + \bm{Y}\cdot \bm{\sigma} |-\rangle)$.  
The Schr\"{o}dinger equation for the $E=0$ states (zero modes) therefore 
amounts to that for the zero modes of 
the untilted (but can be anisotropic) Dirac cones (with zero diagonal elements).
We see that the zero mode is then 
given by 
\begin{equation}
\bm{\alpha}\cdot\bm{\pi} \psi_- = 0 \quad {\rm and} 
\quad  \bm{\alpha}^*\cdot\bm{\pi} \psi_+ = 0.
 \label{zeromode}
\end{equation}
As we shall see in the following, 
the chirality $\pm$ has to be assigned to 
(normalizable) wave functions $\psi_{\pm}$. 
We can also note that the equations  for zero modes apply to spatially 
varying magnetic fields $B(x,y)$ as well, which can even be random.

{\it Aharonov-Casher argument extended to the general chirality ---} 
Is the stability of zero modes inherited by the general chiral symmetry?  
For this we can look at the zero modes for the tilted Dirac cone
 in a spatially varying magnetic field $B(x,y)$, which are therefore  determined by eq.(\ref{zeromode}).  
Here it is convenient to adopt ``principal coordinates" by 
rotating  $\bm{\alpha}$ with an 
orthogonal matrix $T$ (with $\det T = 1)$, so that 
the complex numbers $(z_x,z_y) = (\alpha_X,\alpha_Y)T^{-1}$
become orthogonal with each other on the complex plane.  
It is indeed possible to do this with 
$(z_y/z_x)/|z_y/z_x| = -{\rm sgn}({\rm Im}(\alpha_X\alpha_Y^*))i 
\equiv -i\chi$.  
With the transformed dynamical momentum $\bm{\Pi} \equiv 
T \bm{\pi}$ and the coordinates $\bm{R} \equiv ^t(X,Y) = T \bm{x}$, 
Eq.(\ref{zeromode}) for $\psi_-$ reads 
\begin{eqnarray*}
  \bm{\alpha}\cdot\bm{\pi}  \psi_- 
  =  z_x\lambda^{-1} (\lambda \Pi_X -i\chi \lambda^{-1}\Pi_Y)\psi_-=0,
\end{eqnarray*}
where the ''ellipticity" $\lambda = \sqrt{|z_x|/|z_y|} >0$ is a positive real number. The equation for $\psi_-$ then becomes
$(\lambda \Pi_X - i\chi \lambda^{-1}\Pi_Y)\psi_- = 0$. 
This, along with a similar equation for $\psi_{+}$, reduces to 
\begin{equation}
 (\lambda \Pi_X \pm i \chi\lambda^{-1}\Pi_Y)\psi_\pm = 0.
 \label{zeromode2}
\end{equation}
With these equations, it is straightforward to generalize the analytic 
Aharonov-Casher 
argument\cite{AC}  for the zero modes to the 
present case. 
We use  the ``Coulomb gauge", $\lambda^2\partial_{X}A_{X} + \lambda^{-2}\partial_{Y}A_{Y} =0$, which is 
automatically satisfied if we introduce a ``scalar potential"  $\varphi$ with 
$({A}_{X}, {A}_{Y}) = (-\lambda^{-2} \partial _{Y}\varphi, 
\lambda^2  \partial _{X}\varphi)$.  
Then Eq. (\ref{zeromode2}) simplifies to
$$
 -i\hbar  \bigg[{\cal D}_\pm \mp \chi
  \frac{2\pi}{\phi_0}\bigg( {\cal D}_\pm \varphi\bigg)\bigg]\psi_\pm =0
$$
with ${\cal D}_\pm \equiv (\lambda \partial_{X}\pm i \chi \lambda^{-1}\partial_{Y})$ and  
$\phi_0=h/e$ the flux quantum. 
Putting $\psi_{\pm} = \exp(\pm 2\pi\chi \varphi/\phi_0)\tilde{\psi}_\pm$, we finally obtain 
$$
 {\cal D}_\pm \tilde{\psi}_\pm = (\partial_{\tilde{X}} \pm i\chi \partial_{\tilde{Y}}) 
 \tilde{\psi}_\pm= 0
$$
with $\tilde{\bm{R}} \equiv 
(\tilde{X}, \tilde{Y}) = (\lambda^{-1}X, \lambda Y)$. 
Namely, the function $\tilde{\psi}_\pm$ is an {\it entire function} of $Z_\pm=\tilde{X} \pm i\chi \tilde{Y}$ over the whole complex plane, namely 
a polynomial in $Z_\pm$.  The function $\varphi$ is determined by 
$
B_z = \partial _{X} A_{Y}-\partial _{Y} A_{X}
= (\lambda^2 \partial _{X}^2+\lambda^{-2}\partial _{Y}^2 )\varphi
= ( \partial _{\tilde{X}}^2+\partial _{\tilde{Y}}^2 )\varphi, 
$
which implies that $
\varphi(\tilde{\bm{R}}) = \int d\tilde{\bm{R}'}  
G(\tilde{\bm{R}}-\tilde{\bm{R}'}) B_z(\tilde{\bm{R}'})$
with $
G(\tilde{\bm{R}}) =
 (1/2\pi) \log (r/r_0)$ and $r^2=\tilde{\bm{R}}^2$.
When the magnetic field is nonzero only in a finite region, we have
an asymptotic behavior, 
$\varphi \rightarrow (\Phi/2\pi) \log(r/r_0), $ for $r\to \infty$,
where
$
\Phi = \int d\tilde{\bm{R}} \,B_z
=\int d\bm{R} \, B_z
$
is the total flux.
Then we obtain
$$
\psi_\pm \rightarrow \tilde{\psi}_\pm (r/r_0)^{\pm\chi(\Phi/\phi_0)}
$$
for $r \rightarrow \infty$, in which  $\pm \chi \Phi <0$ is necessary for $\psi_\pm$ to be normalizable.
The normalizability of  $|\psi_\pm|^2$ indicates 
that  the degeneracy of zero modes is ${\Phi}/{\phi_0} $ \cite{AC}.  
This is exactly equal to the total number of 
energy levels in the Landau level, which 
implies a remarkable property that 
{\it no broadening occurs for the $n=0$ Landau level}.

{\it Higher  Landau levels ---} 
 We can also note that 
 the present formulation provides a simple algebraic representation of the 
Landau levels, including higher ones, 
in a uniform  magnetic field $B(x,y)=B>0$,  
where the Schr\"{o}dinger equation reads 
$\bm{\alpha}\cdot \tilde{\bm{\pi}} \psi_- = E\psi_+$ and $\bm{\alpha}^*\cdot \tilde{\bm{\pi}} \psi_+ = E \psi_-$.
Here $ \tilde{\bm{\pi}} = \bm{\pi} - \bm{q}$ 
with a real $\bm{q}$ satisfying 
$\bm{\alpha}\cdot \bm{q} = E\beta$, which amounts to taking the origin to be the center 
 of the cross section (an ellipse) 
of the tilted Dirac cone on a constant-energy plane. 
Since we have $[\tilde{\pi}_x,\tilde{\pi}_y] = [\pi_x,\pi_y]$, an 
annihilation operator $\tilde{a}_E$ satisfying $[\tilde{a}_E, \tilde{a}^{\dagger}_E] = 1$
can be defined as 
$
 \tilde{a}_E = 
\bm{\alpha}\cdot \tilde{\bm{\pi}}/\Lambda_B$ with $\Lambda_B \equiv 
{\sqrt{2{\rm Im} (\alpha_X\alpha_Y^*)\hbar eB}}
$
when $\chi = 1$.  
With a basis $ f_{\ell,E} = (\tilde{a}^{\dagger}_E)^\ell f_0/\sqrt{\ell!}$
for positive integers $\ell$ where $\tilde{a}^{\dagger}_E$ 
acting as the usual raising operator with $\tilde{a}_E f_0 =0$, 
the eigenstate $H\psi_n= E_n \psi_n$ 
with a nonzero energy 
$E_n= {\rm sgn}(n)\sqrt{|n| }\Lambda_B$, found 
in Refs.\cite{MHT,MT,GFMP}, can be algebraically expressed as 
$
 \psi_n  = {\rm sign}(n)|+\rangle  f_{|n|-1,E_n} + | -\rangle f_{|n|,E_n}  
$
for $n=\pm 1,\pm 2,\ldots$.

\begin{figure}[h]
\includegraphics[scale=0.21]{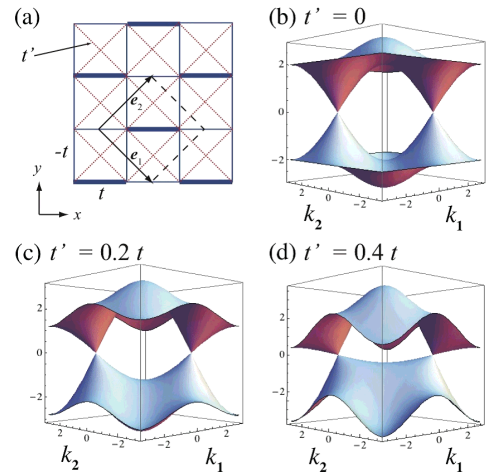}
\caption{(Color Online)
(a) A lattice model possessing tilted Dirac cones.  
Hopping energies are $t$: solid thick lines, $-t$: solid thin lines, 
and  $t'$: dotted lines. A unit cell is indicated by the 
primitive vectors  $\mbox{\boldmath $e$}_1$ and $\mbox{\boldmath $e$}_2$. 
Energy dispersions $E(\bm{k})/t$ for (b) $t'=0$, (c) $t'/t=0.2$ and (d) $t'/t = 0.4$.
\label{fig1}
}
\end{figure}

{\it A lattice model ---} 
In actual materials with general band structures, the 
above argument based on the effective Hamiltonian (\ref{eff-hamiltonian}) and the 
generalized chiral operator $\gamma$
holds only as a low-energy effective model. 
In order to confirm whether the anomalously sharp zero-Landau level 
discussed above appears as well in a lattice model that possesses 
tilted Dirac cones at low energies, 
we consider a  two-dimensional lattice model 
with a  Hamiltonian having nearest ($t$) and 
second neighbor ($t'$) hoppings (Fig.\ref{fig1}) on a square lattice, 
\begin{eqnarray*}
 H &=& \sum_{\bm{r}} -t (c_{\bm{r}+\bm{e}_{y}}^{\dagger} c_{\bm{r}}+{\rm h.c.}) + (-1)^{x+y}t
 (c_{\bm{r}+\bm{e}_{x}}^{\dagger}c_{\bm{r}}+{\rm h.c.}) \\
 &+ &  \sum_{\bm{r}} t' (c_{\bm{r}+\bm{e}_{x}+\bm{e}_{y}}^{\dagger}c_{\bm{r}}+
 c_{\bm{r}+\bm{e}_{x}-\bm{e}_{y}}^{\dagger}c_{\bm{r}}+{\rm h.c.}),
\end{eqnarray*}
where $\bm{r}=(x,y)$ denotes a lattice point in units of the lattice constant, and  $\bm{e}_{x}(\bm{e}_{y})$ the unit vector in 
$x(y)$-direction. 
The primitive vectors for the present lattice system can be chosen as $\mbox{\boldmath $e$}_1 = \bm{e}_{x}-\bm{e}_{y}$ and 
$\mbox{\boldmath $e$}_2 = \bm{e}_{x}+\bm{e}_{y}$ (Fig.\ref{fig1}). 
In the absence of magnetic fields, 
the Hamiltonian in the momentum space becomes
$$
H(\mbox{\boldmath $k$})= \left[\begin{array}{cc}
 2t'(\cos k_1+ \cos k_2)&\Delta (\mbox{\boldmath $k$})\\
\Delta ^* (\mbox{\boldmath $k$}) & 2t'(\cos k_1+ \cos k_2)
\end{array}\right] 
$$
with $\Delta (\mbox{\boldmath $k$}) = -t(-1+e^{ik_1}+e^{ik_1+ik_2}+e^{ik_2})$, where 
$k_1= \mbox{\boldmath $k$}\cdot \mbox{\boldmath $e$}_1$ and $k_2 =
 \mbox{\boldmath $k$}\cdot \mbox{\boldmath $e$}_2$.  
The band dispersion  in the 
first Brillouin zone has a pair of Dirac cones at $(k_1,k_2) = (\pi/2,-\pi/2)$ and $(-\pi/2,\pi/2)$ with 
$E=0$ as long as $|t'/t| < 0.5$, where the tilting becomes stronger 
with $|t'|$ (Fig. \ref{fig1}).  
We can then show that the effective Hamiltonian around these Dirac cones is 
reduced to the Hamiltonian (\ref{eff-hamiltonian}) with ${}^t\bm{W}=(0,\pm 4t')$, ${}^t\bm{X}=(0,2t,0)$ and 
${}^t\bm{Y}=(\mp 2t,0,0)$.
The model reduces  to the $\pi$-flux model\cite{Hatsugai} when $t'=0$ 
(Fig.\ref{fig1}(b)).

We apply a magnetic field to this model to examine the stability of zero modes against the 
disorder in the magnetic field. 
The magnetic field is taken 
into account by the Peierls  substitution $t \rightarrow t e^{-2\pi i\theta(\bm{r})}$, 
$t' \rightarrow t' e^{-2\pi i\theta'(\bm{r})}$ such that the summation of 
the phases $\theta(\bm{r}),\theta'(\bm{r})$ around a loop is equal to the 
encircled magnetic flux in units of the
flux quantum. Here we have adopted 
the string gauge\cite{HIM} to treat smaller magnetic fields.  
Disorder is introduced here as a random component, $\delta \phi(\bm{r})$, 
in the magnetic flux $\phi(\bm{r})= \phi + \delta \phi(\bm{r})$ 
piercing each plaquette,  where 
$\phi$ is the uniform part. 
The random part $\delta\phi(\bm{r})$ is assumed to have 
a gaussian distribution 
with a variance $\sigma$ and a spatial correlation length $\eta$ with 
$\langle \delta \phi(\bm{r}) \delta \phi(\bm{r}')\rangle = \langle \delta\phi^2 
\rangle \exp(-|\bm{r}-\bm{r}'|^2/4\eta)$ \cite{Kawa}.
We have chosen this disorder since it restores, for large enough $\eta$, the 
generalized chiral symmetry of the 
effective Hamiltonian at tilted Dirac cones. 

\begin{figure}
\includegraphics[scale=0.15]{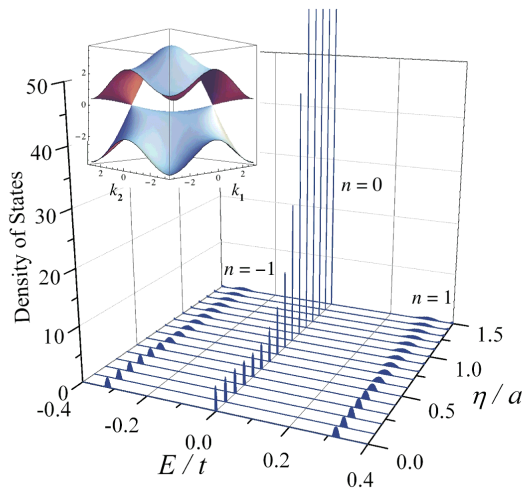}
\caption{(Color Online) Density of states for the 
model with $t'/t = 0.4$ depicted in the inset is plotted against 
the spatial correlation length of the random 
component of the magnetic field $\eta$ 
for a uniform magnetic field $\phi/\phi_0=1/100$ and the 
amplitude of the random magnetic field 
$\sigma/\phi_0 = 0.0029$.  
The result is an average over $5\times10^3$ samples 
with the  system-size 30$a$ by 30$a$. 
\label{fig2}
}
\end{figure}

Figure \ref{fig2} displays the density of states for $t'/t=0.4$, 
for which the tilting is significant  (Fig.\ref{fig1}(d)).  The result 
is obtained 
by the exact diagonalization of a finite system, with an average over $5000$ samples is performed.
It is clearly seen that the $n=0$ Landau level becomes anomalously sharp even in random magnetic fields 
as soon as the correlation length $\eta$ of the random 
flux exceeds the lattice constant $a$, while other Landau levels are 
broadened in a usual fashion. 
This anomaly, appearing only for the  $n=0$  Landau level, suggests that the $n=0$ Landau states are degenerated at $E=0$, 
endorses the stability of zero modes for tilted Dirac cones.

In summary, we have found that the conventional chiral symmetry can be 
extended to a generalized chiral symmetry that encompasses the models having 
tilted Dirac cone, so that  the untilted and tilted Dirac cones can be treated in a unified way.   
The stability of zero modes can be proved under this generalized 
chiral symmetry with an Aharonov-Casher argument. 
We have further shown, numerically for a lattice model, that 
topologically protected zero modes of tilted Dirac fermions 
survive even in random magnetic fields correlated over a few lattice constants.  
These results suggest that the anomaly at 
$n=0$ Landau level can be observed generally in systems with tilted Dirac dispersions.

As a significance of this, we can finally note that 
the existence of the generalized chiral symmetry ($\Delta^2>0$) is equivalent to the 
ellipticity of the Hamiltonian (\ref{eff-hamiltonian}) as a differential operator, under which 
the index theorem \cite{Nakahara} can be applied. 
It is an interesting future problem to extend the notion of the generalized chiral symmetry to a broader class of Dirac-cone systems \cite{watanabe}.

\begin{acknowledgments}
We wish to thank Yoshiyuki Ono and Tomi Ohtsuki
for useful discussions and comments.
The work was supported in part by Grants-in-Aid for Scientific Research,
Nos. 20340098 (YH and HA)
and 22540336(TK) from JSPS and
No. 22014002
on Priority Areas from MEXT for YH. 

\end{acknowledgments}


\vfill
\end{document}